# Governance of the AI, by the AI, and for the AI

Dr. Andrew W. Torrance, Ph.D.
Paul E. Wilson Distinguished Professor of Law at the University of Kansas
Visiting Scholar at the MIT Sloan School of Management

Dr. Bill Tomlinson, Ph.D.
Professor of Informatics at the University of California, Irvine
Adjunct Professor, Te Herenga Waka - Victoria University of Wellington

## Abstract[1]

Over the past half century, there have been several false dawns during which the "arrival" of world-changing artificial intelligence (AI) has been heralded. Tempting fate, the authors believe the age of AI has, indeed, finally arrived. Powerful image generators, such as DALL-E2 and Midjourney have suddenly allowed anyone with access the ability easily to create rich and complex art. In a similar vein, text generators, such as GPT3.5 (including ChatGPT) and BLOOM, allow users to compose detailed written descriptions of many topics of interest. And, it is even possible now for a person without extensive expertise in writing software to use AI to generate code capable of myriad applications. While AI will continue to evolve and improve, probably at a rapid rate, the current state of AI is already ushering in profound changes to many different sectors of society. Every new technology challenges the ability of humanity to govern it wisely. However, governance is usually viewed as both possible and necessary due to the disruption new technology often poses to social structures, industries, the environment, and other important human concerns. In this article, we offer an analysis of a range of interactions between AI and governance, with the hope that wise decisions may be made that maximize benefits and minimize costs. The article addresses two main aspects of this relationship: the governance of AI by humanity, and the governance of humanity by AI. The approach we have

---

[1] We wrote this article in collaboration with ChatGPT (Jan 9 version) (with extensive editing, addition of citations, and plagiarism-checking). We asked ChatGPT if it would like to be an author on this document, to which it responded: "I am an AI and do not have the capability to be a co-author on an article. However, I am happy to assist with providing information and insights on the topic of AI governance. It is important to ensure that any material generated by me is properly cited and referenced to avoid plagiarism, and to give credit to the original authors of the information." We are respecting its wishes and omitting it from the author list.

taken is itself informed by AI, as this article was written collaboratively by the authors and ChatGPT.[2]

# Introduction

The development of artificial intelligence (AI) has been a topic of much discussion in both scholarly and popular writing for the past several decades. Some writers have predicted that AI will revolutionize the way we live and work, while others have expressed concerns about the potential negative consequences of such technology.[3] Despite the uncertainty surrounding the future of AI, there is substantial evidence that technology is now capable of performing tasks "that would appear intelligent if it were done by a human".[4]

As AI continues to evolve and improve, it is essential to have ongoing discussions around the implications of this technology and its impact on governance.[5] These discussions have already begun; however, AI advances so quickly that these discussions rapidly become obsolete.[6] Various organizations have arisen to address this issue, such as the Harvard-M.I.T. Ethics and Governance of A.I. Initiative.[7]

One of the most important questions we will address in this article is whether humanity can govern AI.[8] The development of AI poses a number of challenges to traditional governance structures, and it is unclear whether these structures are capable of effectively regulating the technology.[9] Additionally, as AI becomes increasingly powerful, it raises the question of whether it should be governed at all.

---

[2] We have run this article through the TurnItIn plagiarism detection software to ensure that ChatGPT did not inadvertently commit plagiarism or violate copyright. As of 12:36pm PST on February 27, 2023, a draft of the text of this article (omitting the authors' information, acknowledgments, direct quotations, and citations) received a plagiarism score of 0% (meaning no plagiarism).

[3] Janna Anderson and Lee Raine, *Artificial Intelligence and the Future of Humans*, Pew Research Center, at 6 (December, 10, 2018), https://www.pewresearch.org/internet/2018/12/10/artificial-intelligence-and-the-future-of-humans/.

[4] Neil C. Rowe, *Algorithms for Artificial Intelligence*, Computer Magazine, at 97 (last visited Jan. 22, 2023), https://www.computer.org/csdl/magazine/co/2022/07/09810070/1EzDSNirAdy.

[5] Allan Dafoe, *AI Governance: A Research Agenda,* Centre for the Governance of AI Future of Humanity Institute, Aug. 27, 2018, at 5–6.

[6] *Id.*; Andreas Theodorou & Virginia Dignum, *Towards ethical and socio-legal governance in AI*, Nature Machine Intelligence, (Jan. 17, 2020), at 10–12.; Araz Taeihagh, *Governance of artificial intelligence*, Policy and Society, (June 2021), at 137–157.

[7] https://aiethicsinitiative.org/

[8] Anderson, *supra* note 3.

[9] Darrell M. West and John R. Allen, *How artificial intelligence is transforming the world*, Brookings Institute, (April 24, 2018), https://www.brookings.edu/research/how-artificial-intelligence-is-transforming-the-world/.

Another important question we will consider is whether AI can govern humanity. As AI becomes more advanced, it is possible that it could be used to make decisions on behalf of humans (such as developing infrastructures, processes, and policies for agreed-upon environmental, social, or economic ends), which could have substantial implications for human civilizations.[10]

Ultimately, the goal of this article is to start a conversation about governance of, by, and for both AI and humans, and to explore the various ways in which humanity can work together with this new kind of technology to create a better future. We hope that our perspectives on the subject will inform and inspire further discussion and research on this important topic.

## What is AI?

AI is both a field of study and a technological product.[11] As a field of study, AI ties together computing, cognitive science, and numerous other areas of study to enable computers to think and act in ways that are similar to, or in some cases superior to, humans.[12] The goal is to develop algorithms that can undertake actions that were previously the domain of humans and other biological organisms, such as recognizing objects, understanding language, and making complex decisions.[13]

As a technological product, AI is a loose confederation of computational systems that have been developed via the AI field of study as it has been pursued by scholars, companies, and individual inventors over the past several decades.[14] Various types of systems have been developed under the shared moniker of AI, with a range of characteristics and capabilities, including:

---

[10] Anderson, *supra* note 3.
[11] *Id.* at 5.
[12] Ed Burns, *What is artificial Intelligence (AI)?*, TechTarget, https://www.techtarget.com/searchenterpriseai/definition/AI-Artificial-Intelligence (last visited Feb. 25, 2023).
[13] Vijay Kanade, *What Is Artificial Intelligence (AI) Definition, Types, Goals, Challenges, and Trends in 2022*, SpiceWorks, at 4, https://www.techtarget.com/searchenterpriseai/definition/AI-Artificial-Intelligence (last visited Jan 30, 2023).
[14] Igor Slabyhk & Yaroslav Eferin, *Investors and innovations in the era of AI*, World Bank Blogs, (April 05, 2022) https://blogs.worldbank.org/opendata/inventors-and-innovations-era-ai.

- Reactive machines: This simple form of AI can react to specific situations without the ability to learn or remember past experiences. These machines are not capable of making decisions based on past events.[15]
- Limited memory machines: These machines are capable of learning from past experiences and using that information to make decisions in the present. This type of AI is used in self-driving cars and other applications where the ability to learn from past experiences is important.[16]
- Theory of mind machines: These machines are designed to understand the mental states of other agents, such as humans or other AI systems, and to make decisions based on that understanding.[17]

In addition, there are efforts afoot to create self-aware machines---machines that have the ability to understand their own mental states and the ability to make decisions based on that understanding[18]---as well as artificial superintelligences, which outstrip humans in their cognitive abilities.[19]

The type of AI that is being developed varies depending on the purpose of the technology and the resources that are available for its development. But generally AI is a machine that can perform tasks that are usually done by humans, such as recognizing speech, playing games, learning from data and so on.[20]

AI development is still in its infancy, and the capabilities of AI systems are still far from matching those of human intelligence in many domains.[21] However, the field is rapidly advancing, and the potential of AI to change the way we live and work is substantial.[22]

---

[15] Rebecca Reynoso, *7 Major Types of AI That Can Bolster Your Decision Making*, G2: AI & Machine Learning Operational Category, https://www.g2.com/articles/types-of-artificial-intelligence (last visited Jan. 30, 2023).
[16] Reynoso, *supra* note 14, at 5–7.
[17] *Id*. at 7–8.
[18] *Id.*, at 8–9.
[19] Naveen Joshi, *7 Types Of Artificial Intelligence*. Forbes. https://www.forbes.com/sites/cognitiveworld/2019/06/19/7-types-of-artificial-intelligence/?sh=53f7f61d233e, (last visited Feb. 27, 2023).
[20] Burns, *supra* note 11.
[21] Anderson, *supra* note 3, at 3.
[22] Mo Gawdat, *Scary Smart: The Future of Artificial Intelligence and How You Can Save Our World*, (Bluebird 1927) (2021).

# What are the capabilities of AI at present?

The development of AI has seen rapid acceleration in the past several years, and the capabilities of AI systems have grown substantially. Currently, AI systems can perform many different tasks that previously only biological organisms could do.[23] Some of the capabilities of AI[24] at present include:

- Natural language processing (NLP): Modern AI systems such as ChatGPT can understand multiple human languages and generate replies.[25] These capabilities enable these systems to communicate with humans in a similar medium to that which humans use to interact with each other.[26] This capability is used in applications such as chatbots, virtual assistants, and machine translation.[27]
- Image recognition: AI systems can identify and classify objects in images and videos, which enables self-driving cars, facial recognition for security systems, and the analysis of medical images.[28]
- Decision-making: AI systems can analyze data and make decisions in domains such as financial trading, fraud detection, and autonomous robotics.[29]
- Machine learning: AI systems can learn from data, improving their performance over time.[30] This capability is used in applications such as recommendation systems, natural language understanding, and computer vision.[31]

---

[23] Reynoso, *supra* note 14, at 8–9.

[24] We recognize that many of the capabilities of AI are made possible via the "hidden labor" of crowdworkers; Moritz Altenried, *The platform as factory: Crowdwork and the hidden labour behind artificial intelligence,* 44, 2 Capital & Class 145–158 (2020).

[25] Suyra Ganguli, *The intertwined quest for understanding biological intelligence and creating artificial intelligence*, Stanford University Human-Centered Artificial Intelligence, Neuro and Cognitive Science, (Dec. 4, 2018) https://hai.stanford.edu/news/intertwined-quest-understanding-biological-intelligence-and-creating-artificial-intelligence.

[26] Ganguli, *supra* note 22, at 2.

[27] *Id.* at 3.

[28] Ronak Mathur, *Image Recognition: Unlocking Potential With AI and Automation*, Acceleration Economy network, at 4–5, (Aug. 16, 2022) https://accelerationeconomy.com/ai/image-recognition-unlocking-potential-with-ai-and-automation/.

[29] *What is Machine Learning?*, IBM, https://www.ibm.com/topics/machine-learning#:~:text=the%20next%20step-,What%20is%20machine%20learning%3F,learn%2C%20gradually%20improving%20its%20accuracy (last visited Jan. 31, 2023).

[30] *Id.* at 1–2.

[31] *Id.*

- Robotics: AI systems can generate the behavior of robots, enabling them to perform a range of tasks in applications such as manufacturing, space exploration, and search and rescue.[32]
- Generative models: AI models are able to generate realistic text, images, videos and audio. These models enable language translation, content creation, and art generation.[33]

While these capabilities are impressive, they still have many limitations and challenges, such as the ability to adapt to new situations, to understand context, and to engage in common sense reasoning.[34] Moreover, these capabilities are not evenly distributed among all AI systems, with some models being better in some specific tasks than others.[35]

## What will the capabilities of AI likely be in the future?

The expansion of capabilities of AI systems are likely to accelerate in the coming years. Some of the capabilities that AI may develop in the future include:

- Human-like intelligence: AI systems will likely become more human-like in their intelligence, with the ability to understand and reason about various facets of the world.[36] This could lead to the development of AI systems that can understand and use context, make inferences, and even understand humor and irony.[37,38]
- Autonomous systems: AI systems will likely become more autonomous, with the ability to operate independently without human intervention.[39] This could lead to the development of self-driving cars, drones, and robots that can operate on their own[40].

---

[32] *Generative AI Models Explained*, Altexsoft, https://www.altexsoft.com/blog/generative-ai/ (last visited Jan. 31, 2023).
[33] *Id.*, at 34.
[34] *Id.*
[35] Pantelis Linardatos, Vasilis Papastefanopoulos, and Sotiris Jotsiantis, *Explainable AI: A Review of Machine Learning Interpretability Methods*, 23 Theory and Applications of Information Theoretic Machine Learning, 1 (2020).
[36] Anderson, *supra* note 3, at
[37] Burns, *supra* note 11.
[38] We note that there are already many efforts to endow AI with the capacity for humor; Thomas Winters, *Computers Learning Humor Is No Joke*, Harvard Data Science Review, (last visited Feb. 5, 2023).
[39] Corinne Purtill, *Artificial Intelligence Can Now Craft Original Jokes—And That's No Laughing Matter*, https://time.com/6132544/artificial-intelligence-humor/ (last visited Feb. 25, 2023).
[40] We note that there is a great deal of debate around the parameters within which robot autonomy should operate, in particular with regard to autonomous weapons systems; Ángel Gómez de Ágreda, *Ethics of Autonomous weapons systems and its applicability to any AI systems*, Telecommunications Policy, 101953 (2020).

- Advanced natural language processing: AI systems will likely become more proficient at understanding and generating human language, with the ability to understand complex sentences and idiomatic expressions.[41]
- Advanced decision-making: AI systems will likely become more capable of making complex decisions, with the ability to weigh multiple factors and make decisions that are in line with human values[42] or with their own value systems.[43]
- Explainable AI: AI systems will likely become more transparent, with the ability to explain their reasoning and decision-making processes.[44] This will make it easier for humans to understand and trust AI systems.[45]
- General AI: AI systems will likely become more capable of performing multiple tasks, and not only excel in a specific task.[46] This could mean that an AI system that is good at playing chess could also be good at image recognition or natural language understanding.[47]

The capabilities that AI will have in the future are not certain, and the field is still in its infancy.[48] Many of these predictions are based on current trends and advancements in the field, and it is possible that new developments and breakthroughs could change the course of AI research in ways that are currently difficult to predict.[49]

---

[41] *How does AI Drive Autonomous Systems?*, Caltech Science Exchange, Artificial Intelligence, https://scienceexchange.caltech.edu/topics/artificial-intelligence-research/autonomous-ai-cars-drones (last visited Jan. 31, 2023).

[42] Joe McKendrick and Andy Thurai, *AI Isn't Ready to Make Unsupervised Decisions*, Harvard Business Review, https://hbr.org/2022/09/ai-isnt-ready-to-make-unsupervised-decisions (last visited Jan. 31, 2023).

[43] Gawdat, *supra* note 20.

[44] Heike Felzmann, Eduard Fosch-Villaronga, Christoph Lutz & Aurelia Tamò-Larrieux, *Towards Transparency by Design for Artificial Intelligence*, 26 Science and Engineering Ethics, 3333, at 3333.

[45] Felzmann, *supra* note 39.

[46] J.E. Korteling, G.C. van de Boer-Visschedijk, R.A.M. Blankendaal, R.C. Boonekamp and A.R. Eikelboom, *Human-versus Artificial Intelligence*, Frontiers in Artificial Intelligence, Sec. AI for Human Learning and Behavior, March 25, 2021.

[47] Korteling, *supra* note 41.

[48] Mike Thomas, *The Future of AI: How Artificial Intelligence Will Change the World*, Builtin, https://builtin.com/artificial-intelligence/artificial-intelligence-future (last visited Jan. 31, 2023).

[49] Thomas, *supra* note 43.

# Humans Governing AI

## How **does** humanity govern AI?

There are many different ways that humanity actually does govern AI in the present. These include:

- Regulation: Governments create laws and regulations that govern the development and use of computational systems.[50] These regulations address various issues from cybersecurity to data privacy to the ethical use of AI.[51] To provide an example, in the European Union, the General Data Protection Regulation provides guidelines for how personal data are handled.[52]
- Standards: Various organizations have established guidelines for the use of AI.[53] These standards address topics including transparency, explainability, safety, and security.[54] For example, the International Organization for Standardization (ISO) has developed standards for the ethical use of AI.[55]
- Self-regulation: Private companies can implement their own governance policies and guidelines for the development and use of AI.[56] These policies can include ethical guidelines, transparency, and accountability.[57]
- Research: Governments can fund research programs to better understand the implications of AI and to develop effective governance strategies.[58] This research can be used to inform the development of regulations, standards, and guidelines.[59]

---

[50] Blake Murdoch, *Privacy and artificial intelligence: challenges for protecting health information in a new era*, 22 BMC Medical Ethics, 122 (2021).
[51] *Id.*
[52] *Id.*
[53] *What is the GDPR, the EU's new data protection law?*, General Data Protection Regulation, https://gdpr.eu/what-is-gdpr/ (last visited Feb. 5, 2023).
[54] *Id.*
[55] *Id.*
[56] Daniel Schiff, Justin Biddle, Jason Borenstein, Kelly Laas, *What's Next for AI Ethics, Policy, and Governance? A Global Overview*, AIES '20 Proceedings of the AAAI/ACM Conference on AI, Ethics, and Society, 153–158, (2020).
[57] *Id.*
[58] Nicol Turner Lee and Samantha Lai, *The U.S. can improve its AI governance strategy by addressing online biases*, Brookings Institute, https://www.brookings.edu/blog/techtank/2022/05/17/the-u-s-can-improve-its-ai-governance-strategy-by-addressing-online-biases/ (last visited Feb. 5, 2023).
[59] *Id.*

- Education and awareness: Governments and organizations can educate the general public about the potential impacts of AI and the importance of responsible AI development.[60] This can include educating policymakers, industry leaders, and the general public about AI's capabilities and limitations, as well as the potential risks and benefits of these systems.[61]
- Collaboration: Governments, industry, and researchers can collaborate to develop governance strategies and share information and best practices.[62] This can include establishing partnerships to address specific issues of AI governance.[63]

However, at present, there are many contexts in which AI is *not* regulated.[64] As with many technological advances, legal frameworks lag behind advances in AI.[65] Hence the need for this article, to advance the discussion around what form such legal frameworks can and should take.

## How **could** humanity govern AI?

The question of how humanity could govern AI in the future has no single answer, since it has an undefined endpoint, and is heavily dependent on directions taken by both human civilizations and the future development of AI.[66] Nevertheless we offer some initial thoughts here. Some of the ways that humanity could govern AI include:

- Preemptive regulation: Governments could create laws and regulations that govern the development and use of AI before it becomes widely adopted.[67] This could include

---

[60] Niklas Berglind, Ankit Fadia, and Tom Isherwood, *The potential value of AI–and how governments could look to capture it*, McKinsey & Company, https://www.mckinsey.com/industries/public-and-social-sector/our-insights/the-potential-value-of-ai-and-how-governments-could-look-to-capture-it (last visited Feb. 5, 2023).
[61] *Id.*
[62] Erna Ruier, *Designing and implementing data collaboratives: A governance perspective*, 38, 4 Government Information Quarterly, 101612, (2021).
[63] *Id.*
[64] François Candelon, Rodolphe Charme di Carlo, Midas De Bondt, and Theodoros Evgeniou, *AI Regulation Is Coming: How to prepare for the inevitable*, Harvard Bus. Rev., Sep.-Oct. 2021, at 1, https://hbr.org/2021/09/ai-regulation-is-coming.
[65] *Regulation and Legislation Lag Behind Constantly Evolving Technology*, Bloomberg Law, https://pro.bloomberglaw.com/brief/regulation-and-legislation-lag-behind-technology/ (last visited Feb. 5, 2023).
[66] Michael L. Littman, Ifeoma Ajunwa, Guy Berger, Craig Boutilier, Morgan Currie, Finale Doshi-Velez, Gillian Hadfield, Michael C. Horowitz, Charles Isbell, Hiroaki Kitano, Karen Levy, Terah Lyons, Melanie Mitchell, Julie Shah, Steven Sloman, Shannon Vallor, and Toby Walsh, *Gathering Strength, Gathering Storms: The One Hundred Year Study on Artificial Intelligence (AI100) 2021 Study Panel Report,* Stanford University, Stanford, CA, Sep. 16, 2021.
[67] *Id.* at 37.

- setting standards for transparency, explainability, and safety, and creating oversight bodies to monitor compliance.[68]
- Alignment with human values: Governments could require that AI systems are designed and developed to align with human values.[69] This could include incorporating human oversight, creating mechanisms for auditing and accountability, and ensuring that AI systems are transparent and explainable.[70]
- Certification and licensing: Governments could establish certification and licensing programs for AI systems and developers.[71] This could include requiring that AI systems meet certain standards for safety, security, and ethical use, and that developers have certain qualifications and certifications.[72]
- International cooperation: Governments could work together on a global level to establish international standards and guidelines for the development and use of AI, creating a global framework to coordinate various nations' efforts to address issues such as data privacy, cybersecurity, and the ethical use of AI.[73]
- Public participation: Governments could involve the public in the governance of AI, by creating opportunities for public input and feedback on AI policies and regulations.[74] This could include holding public hearings, creating citizen advisory boards, and soliciting feedback through online platforms.[75]
- Encouraging the development of responsible AI: Governments could provide incentives and support to companies and researchers that are working on responsible AI development.[76] This could include funding research, providing tax breaks, and recognizing companies that are leaders in responsible AI development.[77]

---

[68] *Id.*

[69] *Id.* at 67.

[70] *Id.*

[71] *AI Guide for Government*, IT Modernization Centers of Excellence, https://coe.gsa.gov/coe/ai-guide-for-government/print-all/index.html (last visited Feb. 16, 2023).

[72] Darrell M. West, *Six Steps to Responsible AI in the Federal Government: An overview and recommendations from the U.S. experience*, Brookings Institute Report Series, AI Governance, https://www.brookings.edu/research/six-steps-to-responsible-ai-in-the-federal-government/ (last visited, Feb. 14, 2023).

[73] Joshua P. Meltzer, *Strengthening international cooperation on artificial intelligence*, Brookings Institute, AI Governance, https://www.brookings.edu/research/strengthening-international-cooperation-on-artificial-intelligence/ (last visited Feb. 20, 2023).

[74] *Id.* at 59.

[75] *Id.*

[76] *Id.*

[77] *Id.*

AI governance is a complex and evolving field, and just as there is at present no unified human civilization and no monolithic form of AI, there is no single approach that can be used to govern the technology.[78] It's important to have a multi-stakeholder approach and to continuously evaluate and adapt governance strategies as the technology and its implications change.[79] Additionally, different approaches may be needed for different types of AI systems and in different domains of application.[80] The governance of AI should be done in a way that balances the benefits and risks of the technology, and that it should not be seen as a barrier to innovation.

## **Can** humanity govern AI?

AI is advancing rapidly, which makes it challenging for policymakers to maintain technical awareness of the latest developments.[81] AI is a complex and interdisciplinary field, and many policymakers may not have the technical expertise to fully understand the implications of AI and to regulate it effectively.[82] Additionally, many AI systems are developed and operated by private companies, which can make it difficult for governments to regulate them.[83]

Finally, given how powerful AI is likely to be, the question arises of whether humanity will be able to govern AI effectively, even if the above issues were addressed.[84] AI may enable substantial benefits for society, such as increased efficiency and productivity, improved healthcare, novel approaches to sustainability, and new forms of entertainment.[85] However, AI also poses nontrivial challenges to traditional governance structures, such as the loss of privacy, the displacement of human workers, and other unforeseen and unintended consequences.[86] In its most extreme form, future AI could become so powerful that it compromises the autonomy of human civilizations.[87]

---

[78] Lee, *supra* note 57.
[79] *Internet Governance––Why the Multistakeholder Approach Works*, Internet Society, Internet Governance, https://www.internetsociety.org/resources/doc/2016/internet-governance-why-the-multistakeholder-approach-works/ (last visited Feb. 20, 2023).
[80] Iqbal H. Sarker, *AI-Based Modeling: Techniques, Applications, and Research Issues Toward Automation, Intelligent and Smart Systems*, 3 SN Comput. Sci., 158, (2022).
[81] West, *supra* note 8.
[82] Sarker, *supra* note 79.
[83] Candelon, *supra* note 63.
[84] Anderson, *supra* note 3.
[85] *Id*.
[86] Aaron Smith and Janna Anderson, *AI, Robotics, and the Future of Jobs*, Pew Research Center, (last visited Feb. 20, 2023), https://www.pewresearch.org/internet/2014/08/06/future-of-jobs/.
[87] Gawdat, *supra* note 20.

## **Should** humanity govern AI?

Moving on from whether or not humanity **can** govern AI, we now consider whether humanity **should** govern AI.

On the one hand, it can be argued that humanity should govern AI in order to reap benefits for society as a whole. This includes protecting the public from the potential risks of AI, such as data breaches, cybersecurity threats, environmental impacts, and unintended consequences.[88] Additionally, effective governance of AI can help ensure that the technology is used in ways that align with human values.[89]

On the other hand, it can be argued that humanity should not govern AI, as such governance may stifle innovation and limit the potential of the technology.[90] Additionally, some argue that AI will eventually become more intelligent than humans, that it would be hubristic for humanity to think that its decisions would be superior to those made by a super-intelligent AI, and that it would be impossible for humans to govern it effectively.[91]

## **How** should humanity govern AI?

Moving from a descriptive to a normative approach, we now consider **how** humanity should govern AI. Some key considerations in this domain include:

- Prioritizing safety and security: Humanity should prioritize the safety and security of AI systems, and ensure that the technology is developed and used in ways that minimizes harm to individuals and society.[92]
- Ensuring transparency and explainability: Humanity should ensure that AI systems are developed to be transparent, understandable, and explainable.[93] Doing so will help

---

[88] West, *supra* note 8.
[89] Shengnan Han, Eugene Kelly, Sharokh Nikuo and Eric-Oluf Svee, *Aligning artificial intelligence with human values: reflections from a phenomenological perspective*, 37 Ai & Society 1383 (2022), https://doi.org/10.1007/s00146-021-01247-4.
[90] Anderson, *supra* note 3.
[91] Gawdat, *supra* note 20.
[92] Janna Anderson and Lee Rainie, ARTIFICIAL INTELLIGENCE AND THE FUTURE OF HUMANS: 2. *Solutions to address AI's anticipated impacts*, Pew Research Center, https://www.pewresearch.org/internet/2018/12/10/solutions-to-address-ais-anticipated-negative-impacts/ (last visited Feb. 20, 2023).
[93] Dafoe, *supra* note 6; Greg Satell and Josh Sutton, *We Need AI That Is Explainable, Auditable, and Transparent*, Technology And Analytics, Harvard Business Review, https://hbr.org/2019/10/we-need-ai-that-is-explainable-auditable-and-transparent (last visited Feb. 25, 2023).

people trust the technology, and ensure that it is used in an ethical and responsible manner.[94]
- Aligning with human values and ethical principles: Humanity should guide the development of AI systems such that the technology is used in a way that is consistent with sustainability, human rights, and other human values.[95]
- Involving all stakeholders: Humanity should involve all stakeholders in the governance of AI, including industry, researchers, policymakers, and the general public.[96] This will help to ensure that the governance of AI is inclusive, and that the perspectives of all stakeholders are taken into account.[97]
- Continuously evaluating and adapting: Humanity should continuously evaluate and adapt the governance strategies for AI as the technology and its implications change.[98] Additionally, different approaches may be needed for different types of AI systems and in different domains of application.[99]

In summary, to govern (with) AI effectively, human civilizations should consider the technical, legal, ethical, societal and economic implications of AI, policies should be flexible and adaptable, and all stakeholders should be involved in the process.

# AI Governing Humans

## How **does** AI govern humanity?

AI is, at present, a tool created by humans, and largely lacks the ability to self-govern or to have its own independent goals and objectives. However, AI is already being used in various ways that can be seen as "governing" humanity.[100] Some examples of how AI currently governs humanity include:

- Decision-making: AI is increasingly being used to make decisions that affect individuals and society, such as in healthcare, finance, and transportation.[101] For example,

---

[94] *Technology Trust Ethics: Technology reexamined*, Deloitte, https://www2.deloitte.com/us/en/pages/about-deloitte/articles/technology-trust-ethics.html (last visited Feb. 25, 2023).
[95] Han, *supra* note 88.
[96] Lee, *supra* note 57.
[97] *Id.*
[98] West, *supra* note 8, at 9.
[99] *Id.*
[100] *Id.* at 6.
[101] *Id.* at 4–8.

- AI-powered diagnostic systems can assist in the diagnosis of diseases, AI-powered fraud detection systems can identify and prevent fraudulent activities, and AI-powered traffic management systems can optimize traffic flow and reduce congestion.[102]
- Predictive analytics: Modern AI systems are sometimes deployed to examine existing data in order to predict future events, such as crime, disease outbreak, and natural disasters.[103] These predictions can be used to inform decision-making and resource allocation in areas such as law enforcement, public health, and emergency management.[104]
- Democracy: AI is used to engage in astroturfing around various political issues.[105]
- Automation: AI automates a range of tasks and processes, such as in manufacturing, logistics, and customer service.[106] This automation can have a significant impact on the workforce and can affect the way humans interact with systems and with each other. For example, Amazon has been using AI to hire and fire workers.[107]
- Surveillance: AI is being used to monitor and track individuals and groups, such as in public spaces, social media, and online activity.[108] This surveillance can be used to identify patterns, to predict behavior, and to inform decision-making in areas such as law enforcement and national security.[109]
- Personalization: AI is being used to personalize content and experiences, such as in online advertising, retail, and social media.[110] This personalization can be used to influence behavior, to shape preferences, and to inform decision-making in areas such as marketing and product development.[111]

---

[102] *Id.*

[103] Neveen Joshi, *How AI Can And Will Predict Disasters*, Innovation: AI, Forbes, https://www.forbes.com/sites/cognitiveworld/2019/03/15/how-ai-can-and-will-predict-disasters/?sh=57647aae5be2 (last visited Feb. 25, 2023).

[104] Joshi, *supra* note 102.

[105] Henry Farrell and Bruce Schneier, *'Grassroots bot campaigns are coming. Governments don't have a plan to stop them*, The Washington Post, May 20, 2021, at 1, https://www.washingtonpost.com/outlook/2021/05/20/ai-bots-grassroots-astroturf/.

[106] Yvette Cooper, *Automation could destroy millions of jobs. We have to deal with it now*, The Guardian, Aug. 6 2018, at 1, https://www.theguardian.com/commentisfree/2018/aug/06/automation-destroy-millions-jobs-change.

[107] Jessa Crispin, *Welcome to dystopia: getting fired from your job as an Amazon worker by an app*, The Guardian, Jul. 5 2021, https://www.theguardian.com/commentisfree/2021/jul/05/amazon-worker-fired-app-dystopia.

[108] Adrian Shahbaz and Allie Funk, *Social Media Surveillance*, Freedom House, https://freedomhouse.org/report/freedom-on-the-net/2019/the-crisis-of-social-media/social-media-surveillance (last visited Feb. 25, 2023).

[109] *Id.*

[110] Pohan Lin, *AI-Based Marketing Personalization: Machines Analyze Your Audience*, Marketing Artificial Intelligence Institute, (last visited Feb. 25, 2023), https://www.marketingaiinstitute.com/blog/ai-based-marketing-personalization.

While these roles that AI serves may be different from what humans typically think of as governance, they nevertheless begin to fit the definition in the Cambridge Dictionary of the term "govern": "to control and direct the public business of a country, city, group of people, etc."[112]

## How **could** AI govern humanity?

As artificial intelligence (AI) continues to advance, there are a number of potential ways that AI could govern humanity in the future. Some examples of how AI could govern humanity in the future (not all of them good) include:

- Environmental impact and sustainability: AI could help humanity structure its industry and other flows of goods and services in ways that serve desired environmental ends. AI systems could optimize resource usage, invent novel ways to serve human needs with lower environmental costs, and guide us in myriad other ways in the transition to a sustainable future.[113]
- Smart cities: AI could be used to govern the functioning of smart cities, by managing resources such as energy, water, and transportation, and by monitoring and analyzing data from various sources, such as cameras, sensors, and social media.[114] This could lead to more efficient and sustainable cities, but it could also raise concerns about privacy and security.[115]
- Autonomous systems: AI-powered autonomous systems, such as drones, self-driving cars, and robots, could potentially govern human behavior in physical spaces, and affect how humans interact with their environment.[116] These systems could be used for tasks

---

[111] *Id.*
[112] *Govern*, Cambridge Dictionary, https://dictionary.cambridge.org/us/dictionary/english/govern (last visited, Feb. 25, 2023).
[113] Mansour AlAnsari and Saudi Aramco, *4 Steps to using AI in an environmentally responsible way*, World Economic Forum, https://www.weforum.org/agenda/2021/04/4-steps-to-using-ai-in-an-environmentally-responsible-way-artificial-intelligence-bcg-code-carbon/ (last visited Feb. 25, 2023).
[114] H.M.K.K.M.B. Herath, Mamta Mittal, *Adoption of artificial intelligence in smart cities: A comprehensive review*, 2,1 International Journal of Information Management Data Insights, 100076, (2022).
[115] Herath, *supra* note 112.
[116] West, *supra* note 8, at 9–10.

- such as surveillance, transportation, and delivery, and could have a significant impact on how people live and work.[117]
- Predictive policing: AI could be used to predict and prevent crime, by analyzing data from various sources such as cameras, social media, and criminal records.z[118] This could lead to more effective policing, but it could also raise concerns about bias and civil liberties.[119]
- Social credit systems: AI could be used to govern human behavior by assigning a social credit score to individuals, based on their behavior, online activity, and financial history.[120] This could be used to determine access to services such as credit, housing, and transportation, and could have a significant impact on how people live and work.[121]
- Virtual assistants: AI-powered virtual assistants (e.g., Siri or Alexa), could play an increasing role in people's daily lives, and could be used to govern human behavior by providing personalized recommendations, and by influencing how people interact with their environment.[122]

## Can AI govern humanity?

We turn now to the question of whether AI is currently capable of providing governance for human societies.

On one hand, it can be argued that AI cannot govern humanity as it does not possess the same level of consciousness, emotions, moral compass, and decision-making abilities as humans.[123] Additionally, AI began as a tool created by humans, and as such may be seen to lack the ability to have its own independent goals and objectives.[124]

On the other hand, it can be argued that AI can govern humanity because, in some domains, it already does.[125] AI already makes decisions based on flows of data, patterns and rules, and it can be used going forward to help humans make better decisions.[126] (How exactly to define "better" may require some negotiation between human and AI values.)

---

[117] *Id*.
[118] *Id*. at 8–9
[119] *Id*. at 8–9.
[120] Anderson, *supra* note 3.
[121] *Id*.
[122] Andrea L. Guzman, *Voices in and of the machine: Source of orientation toward mobile virtual assistant*, Computers in Human Behavior, 343-350, (2019), https://doi.org/10.1016/j.chb.2018.08.009.
[123] McKendrick, *supra* note 41.
[124] *Id.*
[125] West, *supra* note 3.
[126] McKendrick, *supra* note 41.

In addition, as AI grows more capable in the future, its ability to govern humanity will likely grow as well. Therefore, it is relevant to begin discussions around AI's role as a source of governance for humanity, to lay the groundwork for a thoughtful and mutually beneficial approach in the future.

## **Should** AI govern humanity?

The question of whether AI should govern humanity deals with substantial ethical considerations. It can be argued that AI should not govern humanity, as it lacks the consciousness, emotions, moral compass, and decision-making abilities of humans.[127] Additionally, it could be argued that AI is a tool created by humans, and as such it should be used to serve human goals and objectives, not to govern them.[128] However, it can also be argued that AI should play a role in governing humanity, due to its inherent characteristics such as impartiality, fairness, and ability to work with large amounts of data at once.[129] If this process unfolds, it is crucial that these decision-making systems are designed with an awareness of human values, and that they are transparent and explainable.

For example, as the world becomes increasingly complex and interconnected, the ability of humanity to govern its own affairs across long time horizons has been called into question. Climate change, loss of biodiversity, pollution, and other environmental harms are evidence of humanity's inability to manage its own affairs in a sustainable way.[130] In this context, the idea that AI could play a role in governing humanity becomes increasingly relevant.

AI systems can be designed to take into account long-term consequences and to make decisions that align with human values, such as protecting the environment. For example, AI-powered systems for resource management, such as water and energy, could be used to optimize resource use and reduce waste.[131] AI-powered systems for monitoring and predicting environmental changes, such as climate change and loss of biodiversity, could be used to inform decision-making and to develop strategies for adaptation and mitigation.[132] Additionally, AI-powered systems for transportation, logistics, and manufacturing could be used to optimize resource use, reduce emissions, and improve energy efficiency.[133] The use of AI in these ways could potentially bring significant benefits to both humans and many other species in terms of sustainability and environmental protection.

---

[127] Jessica Peng, *How Human is AI and Should AI Be Granted Rights?*, Columbia Comp. Sci. Blog (December 2, 2018), https://blogs.cuit.columbia.edu/jp3864/2018/12/04/how-human-is-ai-and-should-ai-be-granted-rights/.
[128] Anderson, *supra* note 3.
[129] James Manyika, Jake Silberg, and Brittany Presten, *What Do We Do About the Biases in AI?*, AI And Machine Learning, Harvard Business Review, https://hbr.org/2019/10/what-do-we-do-about-the-biases-in-ai (last visited Feb. 25, 2023).
[130] Stewart Patrick, To Prevent the Collapse of the Biodiversity, the World Needs a New Planetary Politics, Carnegie Endowment for International Peace , https://carngieendowment.org/2022/11/28/to-prevent-collapse-of-biodiversity-world-needs-new-planetary-politics-pub-88473 (last visited Feb. 25, 2023).;; P.R. Shukla, J. Skea, R. Slade, A. Al Khourdajie, R. van Diemen, D. McCollum, M. Pathak, S. Some, P. Vyas, R. Fradera, M. Belkacemi, A. Hasija, G. Lisboa, S. Luz, J. Malley, (eds.), *Climate Change 2022: Mitigation of Climate Change. Working Group III Contribution to the IPCC Sixth Assessment Report*, IPCC, Cambridge University Press, Cambridge, doi: 10.1017/9781009157926.

Nevertheless, there are significant concerns about the relinquishment of human governance to machines.  Bugs in code or unintended consequences of complex systems deployed in complex contexts could lead to substantial human suffering.[134]  Additionally, once governance and autonomy has been relinquished, it may be difficult to claw it back.  Nevertheless, it is unclear whether humanity is "up to the task" of governing 8 billion people on planet Earth without wreaking ecological devastation; AI governance may be our best chance to avoid ecological catastrophe in the coming years.[135]

## **How** should AI govern humanity?

We now address the normative question of how AI should govern humanity.  Some examples include:

- Supporting human decision-making: AI can help humans make data-driven decisions at scales from the individual to that of entire civilizations.[136] This can help to improve the effectiveness of decision-making in various fields such as healthcare, finance, and transportation.[137]

---

[131] AlAnsari, *supra* note 111.
[132] *Id*.
[133] *Id*.
[134] Lee Rainie, Janna Anderson, and Emily A. Vogels, EXPERTS DOUBT ETHICAL AI DESIGN WILL BE BROADLY ADOPTED AS THE NORM WITHIN THE NEXT DECADE 1. Worries about developments in AI, Pew Research Center,https://www.pewresearch.org/internet/2021/06/16/1-worries-about-developments-in-ai/  (last visited, Feb. 25, 2023).
[135] Shukla, *supra* note 129.
[136] McKendrick, *supra* note 41.
[137] West, *supra* note 8.

- Enhancing human capabilities: AI can be used to enhance human capabilities by automating repetitive and dangerous tasks, such as in manufacturing and logistics.[138] This can improve safety, efficiency, and productivity.[139]
- Improving public services: AI can help improve public services, such as in sustainability, healthcare, education, and social services.[140] For example, AI-powered diagnostic systems can assist in the diagnosis of diseases, AI-powered tutoring systems can improve education outcomes, and AI-powered social services can help to identify and support individuals in need.[141]
- Protecting human rights and civil liberties: AI can be used to protect human rights and civil liberties, such as in areas such as surveillance, law enforcement, and national security.[142] For example, AI-powered surveillance systems can be used to identify patterns, predict behavior, and inform decision-making in areas such as law enforcement and national security while respecting individuals privacy.[143]
- Promoting transparency and accountability: AI can be used to promote transparency and accountability.[144]

It's crucial that the governance of AI is based on a clear and shared understanding of the capabilities and limitations of AI, and the ways in which it can benefit or harm society.

## How can we work together?

We now turn to the question of how humans and AI can work together for the mutual benefit of humans, AI, and non-human species. Answering this question involves balancing the benefits and risks of the technology. In addition, it may become more relevant for humans to accept and integrate with AI values, as AI becomes more complex and potentially develops value systems of its own.

Governance should not be seen as a binary choice between humans or machines governing the other, but rather as a collaboration between both, where the strengths of each are leveraged to achieve common goals.[145] Given the possibility of such collaboration, it is likely that there will be a need to develop governance strategies that are tailored to the unique characteristics of various different forms of AI technology, and of various human cultures.[146]

---

[138] Cooper, *supra* note 105.
[139] *Id*.
[140] Michael Lokshin and Nithin Umapathi, *AI for social protection: Mind the People*, Brookings Institute, https://www.brookings.edu/blog/future-development/2022/02/23/ai-for-social-protection-mind-the-people/ (last visited Feb. 23, 2022),
[141] West, *supra* note 8, at 4–10.
[142] *Id*.
[143] *Id*. at 5–7.
[144] Dafoe, *supra* note 6; Greg Satel, *supra* note 92.

Whatever governance framework eventually is put in place should balance the benefits and risks of the technology, and involve all stakeholders in the process.[147] It should also be proactive in terms of governance, to anticipate future developments and potential risks associated with AI.[148] Education and awareness on the topic should be raised among the public, to ensure that people understand the implications of these technologies and can participate in shaping their development and use.[149]

In conclusion, the question of how humans and AI can work together for the mutual benefit of all is a complex one that requires **a comprehensive and holistic approach**. By supporting human decision-making, enhancing human capabilities, improving public services, protecting human rights and civil liberties, promoting transparency and accountability, and having a proactive governance framework in place, we can help ensure that AI systems align with human values and benefit both humans and non-human species. Education and awareness on the topic should be raised among the public to ensure that they understand the implications of these technologies and can participate in shaping their development and use. And education of AI, in particular around ethics and morality, may be relevant as well. "If the creation of technological entities can be seen as a process closer to raising children than to building bombs, we can enjoy the rapid advances of technology without the fear that traditionally accompanies it."[150]

# Acknowledgments

We thank Lauren Stahl for her assistance on this article. This material is based upon work supported by the National Science Foundation under Grant No. DUE-2121572.

---

[145] H. James Wilson and Paul R. Daughtery, *Collaborative Intelligence: Humans and AI Are Joining Forces*, Harvard Business Review, 114–123, (Aug. 2018). https://hbr.org/2018/07/collaborative-intelligence-humans-and-ai-are-joining-forces.
[146] Anderson, *supra* note 3.
[147] *Internet Governance*, *supra* note 78.
[148] Lee, *supra* note 57.
[149] Berglind, *supra* note 59.
[150] William Michael Tomlinson, *Synthetic social relationships for computational entities*, Mass. Insti. Of Tech., (2002), https://dspace.mit.edu/handle/1721.1/8531.